\documentclass[preprint,12pt]{elsarticle}

\usepackage{amssymb}
\usepackage{amsmath}
\usepackage{tabularx}  % For tables with adjustable width
\usepackage{url}
\usepackage{graphicx}
\usepackage{array}
\usepackage{longtable}
\usepackage[margin=1in]{geometry}
\usepackage{alltt}  % for more control over text formatting
\usepackage[margin=1in]{geometry}
\usepackage{fancyvrb} 
\usepackage{adjustbox}
\usepackage{xcolor}
\usepackage{caption}
\usepackage{multirow}
\usepackage{float}
\usepackage{placeins}
\usepackage{etoolbox}
\patchcmd{\emailauthor}{(#2)}{}{}{}
\patchcmd{\urlauthor}{(#2)}{}{}{}

\begin{document}

\begin{frontmatter}

%% Title, authors and addresses

\title{A Retrieval-Augmented Generation Framework for Academic Literature Navigation in Data Science}

\author[inst1]{Ahmet Yasin Aytar}
\author[inst1]{Kamer Kaya}
\author[inst1]{Kemal Kilic}

\affiliation[inst1]{organization={Sabanci University},%Department and Organization
            addressline={Orhanli-Tuzla}, 
            city={Istanbul},
            postcode={34956}, 
            state={Istanbul},
            country={Turkey}}

\ead{ahmet.aytar@sabanciuniv.edu, 
kamer.kaya@sabanciuniv.edu,
kemal.kilic@sabanciuniv.edu}

\begin{abstract}
In the rapidly evolving field of data science, efficiently navigating the expansive body of academic literature is crucial for informed decision-making and innovation. This paper presents an enhanced Retrieval-Augmented Generation (RAG) application, an artificial intelligence (AI)-based system designed to assist data scientists in accessing precise and contextually relevant academic resources. The AI-powered application integrates advanced techniques, including the GeneRation Of BIbliographic Data (GROBID) technique for extracting bibliographic information, fine-tuned embedding models, semantic chunking, and an abstract-first retrieval method, to significantly improve the relevance and accuracy of the retrieved information. This implementation of AI specifically addresses the challenge of academic literature navigation. A comprehensive evaluation using the Retrieval-Augmented Generation Assessment System (RAGAS) framework demonstrates substantial improvements in key metrics, particularly Context Relevance, underscoring the system's effectiveness in reducing information overload and enhancing decision-making processes. Our findings highlight the potential of this enhanced Retrieval-Augmented Generation system to transform academic exploration within data science, ultimately advancing the workflow of research and innovation in the field.
\end{abstract}


\end{frontmatter}

\section{Introduction}

Like other Generative Artificial Intelligence (GenAI) technologies—such as image, video, and audio generation—Large Language Models (LLMs) that produce text also possess the potential to disrupt both business and social life by transforming how information is processed and decisions are made. However, a significant drawback of LLMs is their current tendency to produce "hallucinations" (i.e., inaccurate or misleading outputs), which can undermine trust in their applications. To address this issue, ongoing research explores Retrieval-Augmented Generation (RAG) as a promising solution to mitigate these inaccuracies. Moreover, across various disciplines, practitioners, and researchers are actively seeking to harness the capabilities of LLMs, particularly as intelligent assistants to enhance decision-making processes. RAG can potentially improve the effectiveness of intelligent assistants by integrating relevant information from external sources, allowing for more accurate and contextually appropriate responses. Data science, as a critical field in the realm of digital transformation that relies on data-driven insights, is no exception; it stands to benefit from advancements in LLM technology while addressing its inherent weaknesses.

Data science is an exploratory process that necessitates the formulation and validation of hypotheses using historical datasets. The models developed by data scientists essentially represent these hypotheses, and creating effective models relies on the formulation of pertinent questions. Therefore, data scientists benefit from exposure to various alternatives, and academic literature serves as a vital resource that enhances their creative thinking abilities. However, as data science continuously evolves, it generates an overwhelming volume of research, making the efficient access to and utilization of academic literature increasingly challenging. This surge in available information complicates the ability of researchers and practitioners to identify and integrate relevant insights into their work, highlighting the need for effective tools and strategies to navigate this expanding landscape.

To mitigate this issue, in this research, we have developed a RAG application specifically designed to aid data scientists in navigating academic resources and obtaining guidance tailored to their unique needs. This application encompasses a broad spectrum of data science subfields, including machine learning, deep learning, time series modeling, synthetic data generation, reinforcement learning, data preprocessing, feature engineering, and evaluation. The value of our RAG application is evident in its ability to provide precise and contextually relevant responses to specific queries posed by data scientists. This tool not only conserves time but also improves the accuracy of decisions made throughout the data science workflow by integrating advanced techniques for retrieving and embedding pertinent academic insights into a LLM. Consequently, the application enables users to make well-informed decisions regarding methodologies and techniques, thereby contributing to the development of more effective and innovative solutions for their projects.

Our objective with this application is to systematically combine and assess the performance of a range of established techniques within the RAG framework, reporting their effectiveness within a particular domain to generate insights valuable to researchers. By conducting a comprehensive evaluation of these methods in a real-world setting, we aim to facilitate the broader adoption of these approaches in both academic and industrial contexts.

Our RAG application responds to a significant need within the data science community for accessible and relevant academic guidance. Unlike traditional literature review methods, which can be time-consuming and overwhelming due to the vast amount of available information, our application facilitates the process by extracting and embedding only the most pertinent sections of articles into the LLM. This approach accelerates decision-making and ensures that the information provided is both current and directly relevant to the user’s query, thereby addressing the prevalent issue of information overload in data science research.

The primary innovation of our RAG application lies in its 
\textit{five-stage enhancement process}, which substantially improves the traditional RAG pipeline. Initially, we utilize GROBID (GeneRation Of BIbliographic Data) to clean and structure article data before constructing the vector database, ensuring the accuracy and organization of the information \cite{Lopez2024}. Next, we implement a specialized fine-tuning process using academic textbooks, which enhances the embedding model's capacity to interpret and process complex queries. Furthermore, our application employs semantic chunking rather than traditional recursive chunking which ensures that text is divided into meaningful units and results in a more coherent and relevant vector database \cite{Kamradt2024}. We also adopt an abstract-first method, prioritizing searches within a separate vector database of article abstracts, thereby streamlining the retrieval process. Finally, advanced prompting techniques are employed to optimize the LLM’s responses, leading to more accurate and contextually appropriate answers. Llamaindex and LangChain were instrumental in creating fine-tuning data and RAG pipelines, respectively \cite{Liu2022, Chase2022}. Additionally, OpenAI's GPT-4o model was employed as the LLM throughout the study, chosen for its cost-effectiveness and efficiency.

To validate the effectiveness of our application, we employed the RAGAS (Retrieval Augmented Generation for Academic Search) framework and conducted tests with 50 sample questions covering various domains and problem types \cite{Es2023}. This rigorous testing process confirmed the application’s capability to deliver high-quality, relevant responses, highlighting its potential as a useful tool for data scientists. By integrating these advanced techniques, our application significantly enhances the traditional RAG pipeline, providing data scientists with a practical tool that simplifies academic exploration and supports informed decision-making.

To validate the effectiveness of our application, we employed the RAGAS (Retrieval Augmented Generation for Academic Search) framework and conducted tests with 50 sample questions covering various domains and problem types \cite{Es2023}. This rigorous testing process confirmed the application’s capability to deliver high-quality, relevant responses, highlighting its potential as a useful tool for data scientists. Although all metrics in the RAGAS framework—Context Relevance, Faithfulness, and Answer Relevance—are important, our primary focus was on improving Context Relevance. This metric serves as the main indicator of the system's ability to retrieve pertinent academic content, which aligns with the central objective of this study: enhancing the retrieval process for relevant information in academic literature for data scientists. The other metrics, such as Faithfulness and Answer Relevance, assess the quality of answers generated using the retrieved context, but since the language model used for answer generation remained unchanged, these metrics are less indicative of improvements brought by the fine-tuning process and other retrieval enhancements. Therefore, Context Relevance serves as the most appropriate measure to demonstrate advancements in retrieval capabilities across different configurations.

This paper makes two primary contributions. First, it introduces an enhanced RAG architecture and systematically evaluates the impact of each component on performance using key metrics within the RAGAS framework, based on a specially prepared test set of relevant questions. Second, it provides a focused analysis of the fine-tuning process, characterized by its experimental flexibility and diverse potential approaches. This analysis involves exploring various alternative configurations by altering data sources and preprocessing techniques to allow for a detailed assessment of how different fine-tuning strategies influence the metrics. The most effective configurations are then integrated into the overall evaluation of the proposed architecture, which contributes to a comprehensive understanding of each component's role in system performance.

The remainder of the paper is organized as follows: The next section reviews the literature on RAG and outlines the components of the five-stage enhancement process, highlighting existing research gaps. In Section 3, we introduce the enhanced RAG architecture and the metrics used to evaluate the performance of various alternatives. Section 4 presents a detailed overview of the key findings from the experimental analysis. In the following section, we provide an in-depth discussion and analysis of these findings. Finally, the paper concludes with final thoughts and a discussion of potential avenues for future research in Section 6.

\section{Literature Review}
Academic guidance is an essential element in the work of a data scientist, providing a framework for grasping existing research and directing new, creative project directions. Engaging with scholarly articles allows data scientists to identify established methodologies, discern gaps in current research, and draw connections between their specific challenges and those addressed in prior studies. This process is essential for both refining research questions and selecting or adjusting methods for data analysis and model development. As noted by \cite{Mysore2023}, data scientists heavily depend on the literature to assess what has been accomplished, determine whether similar problems have been addressed, and find solutions to related challenges.

In recent years, the volume of literature on AI and machine learning has expanded rapidly, with publications doubling approximately every 23 months as of 2022 \cite{Krenn2022}. This continual growth has resulted in an overwhelming amount of information for data scientists to sift through. The difficulty of pinpointing the most relevant and high-quality articles amid the vast number of publications adds complexity to the task. Consequently, data scientists often rely on keyword searches or similar techniques, which may be inadequate due to the difficulty in crafting effective search queries. They may struggle to identify the appropriate terms or strategies to locate relevant articles efficiently and promptly, leading to potential gaps in their literature reviews \cite{Mysore2023}.

A promising approach to addressing the challenge of navigating the extensive literature in data science is Retrieval-Augmented Generation (RAG) \cite{Lewis2020}. RAG's capacity to access external knowledge bases makes it a valuable tool for managing and comprehending the vast literature in the field. The integration of Large Language Models (LLMs) within the RAG pipeline further enhances this capability by making the process of understanding complex academic papers and their core contributions more manageable. By retrieving contextually relevant information and summarizing it effectively, RAG assists data scientists in quickly grasping the key contributions of numerous papers, thus mitigating the burden of information overload. RAG models have demonstrated considerable promise in tasks such as question-answering and dialogue systems, where the ability to dynamically retrieve pertinent information from external sources is essential \cite{Izacard2020}. This capability is particularly valuable in the fast-growing field of data science, where synthesizing and analyzing an expanding body of literature is critical.

RAG has been adopted across various industries, including healthcare, legal, and education to enhance domain-specific applications and address the challenges of these fields. In healthcare, RAG models support clinicians by providing evidence-based recommendations from up-to-date medical literature and databases, enhancing diagnostic accuracy and treatment planning. For example, the MKRAG framework has improved medical question answering by integrating medical facts from external knowledge bases into the generation process, thereby increasing the accuracy of responses and aiding clinicians in making informed decisions \cite{Shi2023a}. In the legal field, one method to improve question-answering systems involves the integration of RAG models with Case-Based Reasoning (CBR). This CBR-RAG approach enhances LLM queries with contextually relevant legal cases, improving the retrieval of pertinent precedents and statutes. This method is among the strategies developed to provide more accurate and contextually relevant answers to complex legal queries, thereby increasing the efficiency and accuracy of legal research \cite{Wiratunga2024}. Additionally, systems like LegalAsst utilize AI-powered tools for precise legal case retrieval, incorporating structured representations and decision-tree-based judgments to streamline court procedures, thus enhancing legal assistance and court productivity through improved retrieval accuracy of similar cases and relevant legislation \cite{Han2024}. In education, RAG models are utilized to improve math question-answering by incorporating vetted external content, such as scripts from high-quality open-source textbooks, into LLM prompts. This method has been shown to improve response quality, assisting students and educators in navigating complex problems with contextually relevant support, although designers must balance the trade-offs between aligning responses with educational resources and meeting student preferences \cite{Levonian2023}. 

RAG applications have significantly streamlined the academic literature review process. For example, KNIMEZoBot \cite{Alshammari2023} integrates Zotero with KNIME and OpenAI's language models to automate citation management and literature generation, reducing the time and effort required for comprehensive reviews. LitLLM \cite{Agarwal2024} further enhances this process by synthesizing large volumes of research into concise, field-specific summaries, making it an invaluable tool for scientific researchers. In the biomedical field, RefAI \cite{Li2024} employs a GPT-powered RAG system to manage complex terminologies and research details with precision. Collectively, these tools illustrate the growing influence of RAG technology in transforming literature reviews, offering tailored and efficient solutions across diverse academic disciplines.

Despite the advancements offered by RAG applications, substantial challenges remain, particularly regarding information overload and contextual relevance. For instance, current models often retrieve passages indiscriminately, without regard to whether they are factually grounded or contextually appropriate, which can result in low-quality outputs containing irrelevant or off-topic information \cite{Shi2023b}. Furthermore, the accuracy and reliability of the generated content are not guaranteed, as these models are not always explicitly trained to leverage facts from the retrieved passages, leading to potential inconsistencies \cite{Gao2023}. Additionally, handling complex and domain-specific queries remains a significant issue, as RAG models often require fine-tuning to improve their accuracy and effectiveness in specialized fields \cite{Siriwardhana2023}. Moreover, obtaining a high-quality retrieval model that effectively fetches relevant documents is an ongoing challenge that impacts the overall performance of these systems \cite{Arora2023}.

To address the limitations identified in traditional Retrieval-Augmented Generation (RAG) models, we propose a comprehensive five-stage enhancement process that significantly improves the performance, accuracy, and relevance of RAG applications. This process is designed to handle the complexities of domain-specific queries, ensuring that the information retrieved is not only pertinent but also aligned with the user's context and intent. Each stage of our approach contributes to refining different aspects of the RAG system, from data cleaning and embedding fine-tuning to semantic chunking, abstract-first retrieval, and advanced prompting techniques. Collectively, these innovations have the potential to significantly reduce information overload and improve the contextual relevance of responses, thereby providing a robust solution to the common challenges encountered by traditional RAG models.

A key component of our approach is the use of GROBID \cite{Lopez2024}, an open-source tool renowned for its accuracy in reference parsing and metadata extraction. GROBID is utilized in our application to structure and clean data from PDFs, ensuring the information is well-organized before being embedded into our vector database. A study evaluating multiple reference parsing tools demonstrated that GROBID outperformed its peers, achieving the highest F1 score of 0.89 among the ten tools tested \cite{Tkaczyk2018}. This high level of precision is crucial for maintaining the quality of metadata and ensuring that a RAG system delivers accurate and contextually appropriate responses. Moreover, GROBID's success in various contexts, such as automated metadata extraction in the CERN Document Server and citation generation in CiteAssist, further underscores its versatility and reliability \cite{Kaesberg2024, Boyd2015}.

In the second stage of our approach, we emphasize the fine-tuning of the embedding model to enhance its capability to accurately retrieve domain-specific information, a critical step in improving the overall effectiveness of our RAG application. This fine-tuning process draws on methodologies from recent studies, which highlight the importance of adapting models to the specific contexts in which they are applied. For instance, \cite{Pu2024} demonstrated how domain-specific fine-tuning significantly improved retrieval accuracy for technical documentation in Electronic Design Automation (EDA) tools by benchmarking the performance of various models. Similarly, \cite{Nguyen2024} combined fine-tuning with iterative reasoning in a Q\&A system, allowing for progressively refined answers, especially in complex scenarios. Natural Language Inference (NLI) to fine-tune embedding models is employed in \cite{Dusek2023}, enhancing their ability to understand nuanced domain-specific language, thereby improving the relevance and accuracy of retrieved content. Therefore, by incorporating these embedding fine-tuning techniques, we hypothesized that the RAG application would be better equipped to handle specialized and intricate queries common in data science workflows, ensuring that the information provided is both precise and contextually appropriate.

In the third stage of our approach, we apply semantic chunking to improve the coherence and relevance of the data used in our RAG application \cite{Kamradt2024}. This technique enables breaking down complex information into semantically meaningful units, which are then more effectively mapped to relevant knowledge sources. The application of semantic chunking has demonstrated significant usefulness in specific fields. For example, \cite{Wu2024} applied semantic chunking to organize medical knowledge into graph-based structures, ensuring the safe and contextually appropriate retrieval of medical information—a critical feature in healthcare where accuracy is essential. Additionally, \cite{WuMooney2022} utilized semantic chunking in Visual Question Answering (VQA), where it was used to decompose complex visual questions into simpler, semantically coherent parts. This approach significantly enhanced the model's ability to interpret and respond accurately by linking these chunks to external knowledge sources. Consequently in the enhanced RAG architecture we considered integrating a similar semantic chunking technique to maintain high levels of contextual accuracy and relevance, ensuring that the information retrieved and presented is both coherent and directly applicable to the user’s query.

In the fourth stage of our enhancement process, we introduce an "abstract-first" method for RAG applications, which prioritizes searches within a separate vector database of article abstracts. This approach is intended to enhance the retrieval process by concentrating on the most concise and relevant sections of academic papers, which are the abstracts. The rationale behind this method is that abstracts typically provide a clear and succinct summary of the research, capturing the core findings and methodologies. Consequently, many existing datasets used for Named Entity Recognition (NER) models, which are essential for NLP tasks such as information extraction, concentrate solely on abstracts and preselected texts \cite{Otto2023}. This strategy mirrors the natural tendency of researchers to first read abstracts when assessing the relevance of a paper \cite{Sambar2024}. By targeting these sections first, a RAG application can quickly and efficiently determine the relevance of a document, thereby improving retrieval efficiency and reducing computational load. This approach is particularly advantageous in fields where quick access to respective information is critical, addressing the common challenge of exploring large volumes of full-text content to identify key insights. The abstract-first method thus potentially represents a significant advancement in the design of RAG systems, offering a sophisticated alternative to traditional full-text search strategies.

In the fifth stage of our enhancement process, we implement advanced prompting techniques to optimize the responses generated by our RAG application. This stage involves incorporating strategies such as tip-offering and emotional prompting, which have been shown to enhance the performance and contextual relevance of Large Language Models (LLMs). For example, \cite{Salinas2024} demonstrated that even minor adjustments to prompts, such as offering tips or slight rephrasing, can lead to substantial improvements in the precision and relevance of LLM outputs. This underscores the importance of carefully crafted prompts in refining the responses generated by RAG systems. Similarly, \cite{Li2023} showed that emotional prompting, where prompts are infused with emotional cues, can improve the LLM's ability to produce more empathetic and contextually rich responses, which is particularly valuable in applications requiring a nuanced understanding of user needs, such as mental health support or customer service. By integrating these advanced prompting techniques into the RAG model, we aim to improve the performance of the LLM while also offering prompts that direct the model to produce more precise and contextually relevant responses, ultimately enhancing the utility and effectiveness of our application.

\section{Methods}
In this section, we will first present the original RAG approach and provide the details of the enhanced RAG architecture we have developed in this research. Additionally, we will explain the experimental design and present the performance metrics used to evaluate the effectiveness of the alternatives.

\subsection{Overview of RAG Architecture and Data Preparation}
\begin{figure}
    \centering
    \includegraphics[width=0.9\linewidth]{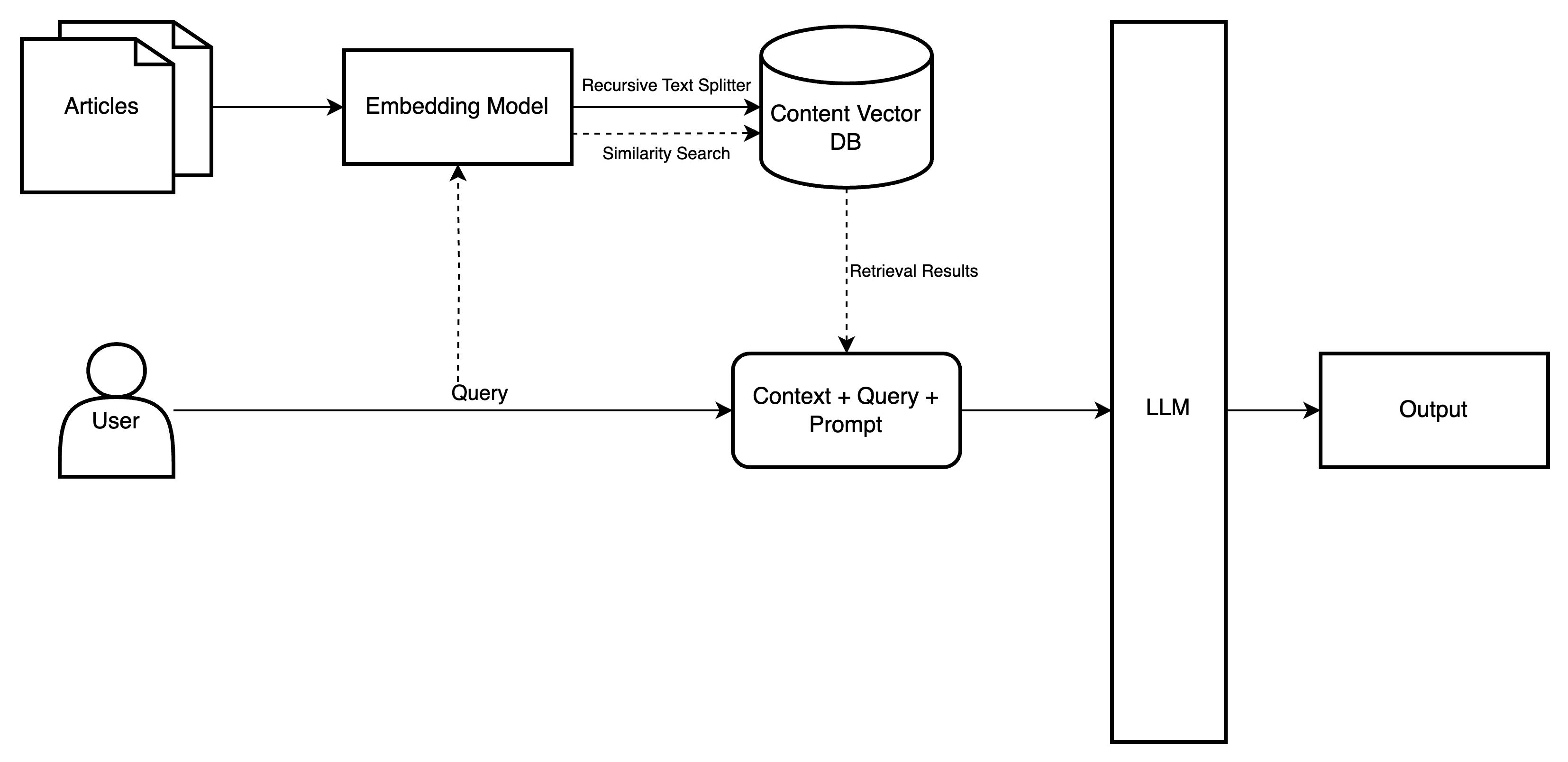}
    \captionsetup{font=small} % Set font size for this caption
    \caption{Original RAG Architecture}
    \label{fig:Original_RAG}
\end{figure}

\begin{figure}
    \centering
    \includegraphics[width=0.9\linewidth]{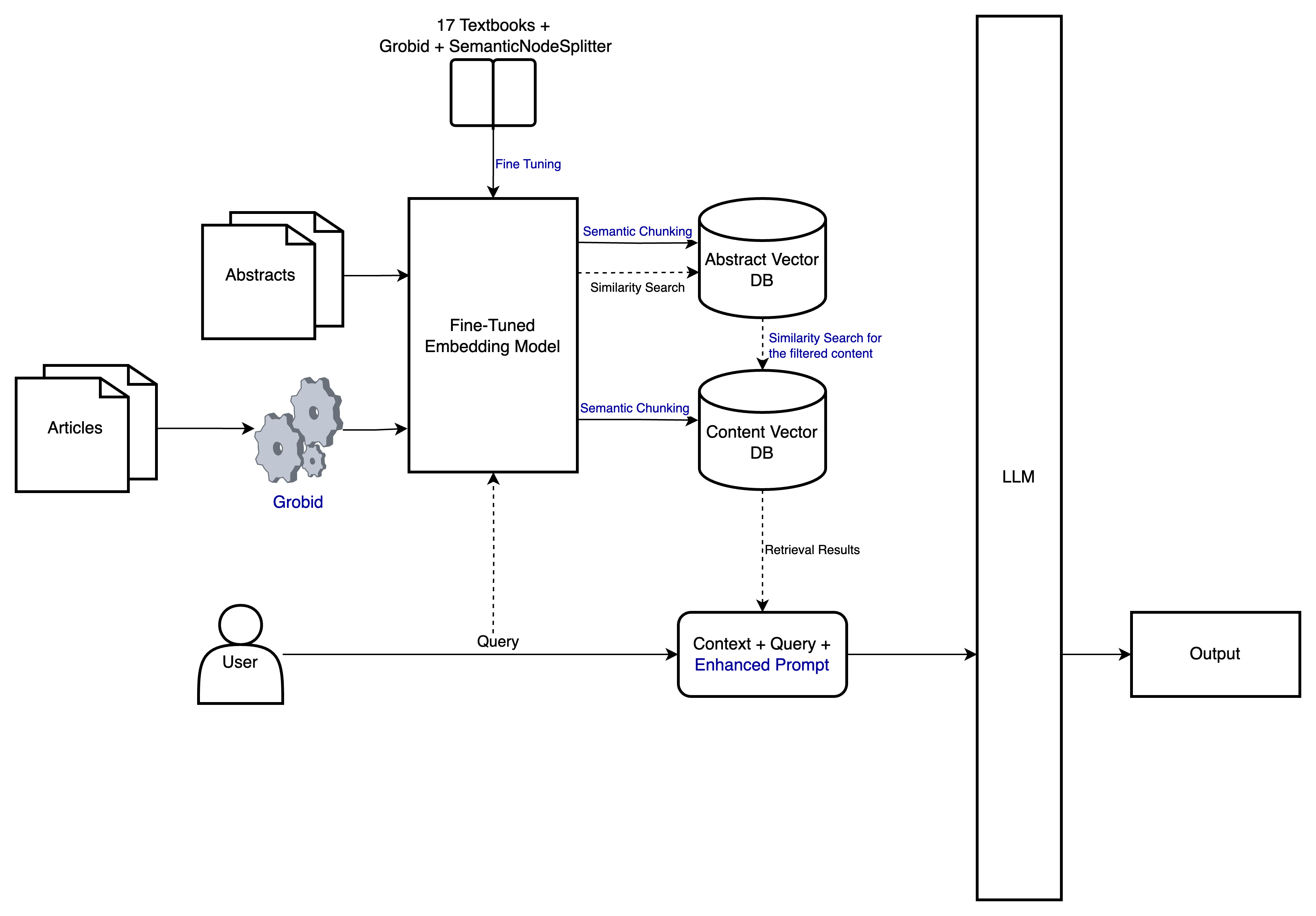}
    \captionsetup{font=small} % Set font size for this caption
    \caption{Our Enhanced RAG Architecture}
    \label{fig:Enhanced_RAG}
\end{figure}

In the progression from an original RAG architecture to our enhanced version, significant modifications were made to improve the system’s accuracy, relevance, and efficiency. Figure \ref{fig:Original_RAG} illustrates the basic RAG architecture, where a user’s query is first embedded and then used to search a vector database for relevant document chunks. These retrieved chunks, along with the original query, are then fed into a language model to generate the final output. In contrast, Figure \ref{fig:Enhanced_RAG} showcases our enhanced RAG architecture, where we have introduced several critical innovations, highlighted in blue, to address the limitations of the classical approach. These enhancements include advanced data cleaning, fine-tuned embedding models, semantic chunking, a novel abstract-first search strategy, and improved prompting techniques. The details of these improvements are outlined below.

To thoroughly evaluate our enhanced RAG application, we developed a comprehensive set of 50 questions covering various sectors such as telecommunications, agriculture, and others, reflecting the diverse challenges encountered in data science. These questions were meticulously crafted according to the CRISP-DM (Cross-Industry Standard Process for Data Mining) methodology \cite{ncr1999crisp}, a widely recognized framework in the data science community. The question set spans the three major stages of the CRISP-DM framework: Data Preparation, Modeling, and Evaluation, with 10 questions dedicated to Data Preparation, 35 to Modeling, and 5 to Evaluation.

Within the Data Preparation stage, we targeted essential areas such as feature selection (3 questions), outlier elimination (2 questions), and dimension reduction (5 questions). The Modeling stage, which often demands the most academic guidance, included questions on synthetic data generation (3 questions), classification (5 questions), regression (5 questions), clustering (5 questions), reinforcement learning (2 questions), image recognition (5 questions), natural language processing (5 questions), and time series analysis (5 questions). Lastly, the Evaluation stage comprised 5 questions focused on assessing model performance. Each question was designed to address specific subdomains and techniques, ensuring a rigorous assessment of the RAG application across multiple aspects of data science. The complete list of questions is provided in Appendix A.

For instance, one of our questions is: "As a data scientist in the financial sector tasked with predicting stock prices, I have used traditional feature selection methods like backward elimination. Which advanced feature selection technique should I consider to enhance model performance?" This format ensured that the questions were realistic and reflective of the actual challenges faced by professionals in the field.

In our analysis, arXiv API was employed to retrieve relevant articles necessary for constructing a vector database. We used arXiv \cite{Wong2017} since it is a prominent platform that allows researchers to share their scientific and academic papers before undergoing the peer-review process \cite{Khouya2024}. For the analysis, each test question was first translated into a detailed search query, compatible with the arXiv API. This conversion process was facilitated by GPT-4o using a specific prompt designed to condense each question into a targeted search query: "Transform the following detailed inquiry into a concise and focused research query suitable for academic search databases. The summary should highlight the main topic, applicable domain, and specific requirements or goals in a maximum of 10 words." This approach ensured that each search term accurately represented the specific research objective associated with each question. Approximately 100 relevant articles were selected for each question, resulting in a dataset comprising roughly 5,000 articles. The content of the selected articles was downloaded as PDF files and subsequently used in the data-cleaning step. In parallel, the metadata for each article —such as publication date, title, authors, and abstracts— retrieved from the arXiv API, was also utilized. Following data cleaning, both the cleaned article content and the associated metadata were systematically integrated into the vector database. This process ensured that the database facilitated structured retrieval by academic standards.

\subsection{Data Cleaning using GROBID}

Upon acquiring the articles, we initiated our five-stage enhancement process, starting with data cleaning. Since the articles retrieved from the arXiv database were in PDF format, it was necessary to parse and convert them into text while eliminating relatively less relevant data such as figures, symbols, and extraneous spaces. To accomplish this, we used GROBID (GeneRation Of BIbliographic Data), a machine-learning-based tool that excels in extracting, parsing, and restructuring raw documents into clean, structured text \cite{Lopez2024} as previously mentioned. This step was critical because the original data contained numerous non-text elements that could degrade the quality of the vector database. By ensuring that the data is well-organized and free from noise, GROBID seeks to establish a basis for producing accurate and significant vector embeddings.

\subsection{Fine-tuning}

The second stage focused on fine-tuning the embedding model to improve its ability to capture domain-specific nuances to ensure that both the articles and queries were embedded in a manner that accurately reflected the complexities of data science and academic literature. Given that the embedding model is essential for converting both queries and articles into vectors for storage in the vector database, the model needed to be proficient in understanding complex academic language. We conducted a series of experiments to fine-tune the "sentence-transformers/all-mpnet-base-v2" model using a range of textbooks and various techniques \cite{huggingface_mpnet_base}. These experiments included:

\begin{enumerate}
    \item \textbf{Experiment 1:} Fine-tuning with 5 textbooks.
    \item \textbf{Experiment 2:} Fine-tuning with 17 textbooks.
    \item \textbf{Experiment 3:} Fine-tuning with 5 textbooks, incorporating GROBID for additional data cleaning.
    \item \textbf{Experiment 4:} Fine-tuning with 17 textbooks, incorporating GROBID for additional data cleaning.
    \item \textbf{Experiment 5:} Combining 17 textbooks, GROBID, and Semantic NodeSplitter.
\end{enumerate}

The textbooks, listed in Appendix B and Appendix C, cover the same topics as our question set, ensuring that the fine-tuned models were well-aligned with the content they were intended to process. The fine-tuning process involved converting each textbook into nodes —coherent segments of text—and then generating question-answer pairs for each node using the OpenAI API. GROBID, which specialized in processing academic content, was also employed to clean the fine-tuning dataset before creating these Q-A pairs. GROBID's application to the textbooks resulted in well-prepared text for subsequent stages. As emphasized by \citep{Gunasekar2023}, the importance of high-quality data in the context of fine-tuning cannot be overstated. This insight drove our decision to use GROBID for data cleaning, ensuring the integrity and relevance of the academic content used in our experiments \citep{Lopez2024}. Additionally, by incorporating the Semantic Node Splitter, which operates similarly to semantic chunking, we were able to segment nodes based on the semantic similarity of sentences rather than arbitrary text lengths \citep{Kamradt2024}. This approach allowed us to create nodes that were more semantically coherent instead of using preset token size to create nodes, which resulted in more meaningful Q-A pairs and a fine-tuned model better aligned with the complexities of academic content.

The number of Q-A pairs for each experiment is as follows: Experiment 1 produced 4,819 pairs, Experiment 2 resulted in 15,326 pairs, Experiment 3 generated 1,600 pairs, Experiment 4 created 4,460 pairs, and Experiment 5 yielded 4,132 pairs (altogether 30,337 pairs). As the number of textbooks increased between Experiments 1 and 2, the number of Q-A pairs grew substantially due to the greater volume of available data. However, when GROBID was applied in Experiments 3 and 4, a significant reduction in Q-A pairs was observed, as GROBID systematically removes extraneous content and focuses on the core textual material relevant for model fine-tuning. When the Semantic Node Splitter was introduced in Experiment 5, the number of Q-A pairs remained relatively consistent by adjusting the similarity threshold during node splitting to provide the optimal balance between quantity and the meaningfulness of the Q-A pairs. Since there is no established standard for the optimal number of Q-A pairs required for fine-tuning an embedding model, these varied experiments are designed to generate datasets of different sizes and structures for fine-tuning. This variability allows us to explore how different data formats influence the model's performance and gain insights into the relationship between dataset size, content quality, and the resulting fine-tuned embeddings \cite{Lopez2024}.

\subsection{Semantic Chunking}

In the third stage, we employed semantic chunking to build the vector database \cite{Kamradt2024}. Unlike traditional methods that rely on recursive chunking based on a fixed token size, semantic chunking preserves the contextual integrity of the text by grouping sentences that are semantically related. The process begins by splitting the text into individual sentences, which are then converted into vector embeddings. By calculating the cosine similarity between these embeddings, we grouped sentences into semantically coherent chunks. A similarity threshold was needed to be chosen to separate the chunks so that they were neither too fragmented nor overly broad. Also, achieving a comparable word count was crucial to ensure that the metrics were fair and directly comparable across different experiments. Through careful adjustment of the similarity threshold, semantically coherent sentence clusters were formed while preserving a comparable word count to other experimental contexts, thereby ensuring the equitable comparability of metrics. This approach in general was particularly important because it directly influenced the performance of our RAG application in terms of both the relevance and coherence of the retrieved information. These findings are consistent with previous research suggesting that relationships between words within a sentence often convey more about the underlying semantic content than the individual words themselves \cite{Ruas2020}.

\subsection{Abstract-first Method}

Our fourth stage introduced an innovative abstract-first search strategy, designed to simplify the retrieval process by initially searching a vector database composed solely of article abstracts. Abstracts, being concise summaries of research findings and methodologies, offer a quick and efficient means of assessing the relevance of a document. By embedding the query and searching within this abstract database first, we could filter out irrelevant articles early in the process. The titles of \textit{the top 100 most relevant articles} were then used to conduct a more focused search in the larger vector database containing the full articles. As detailed in the experimental analysis we will discuss later, this method significantly improved search efficiency and accuracy, reducing computational load and speeding up retrieval times.

\subsection{Enhanced Prompting Technique}

In the final stage, we optimized the interaction between the LLM and the retrieved context by experimenting with various prompting techniques. Although the default prompt provided by LangChain served as a solid baseline ("You are an assistant for question-answering tasks. Use the following pieces of the retrieved context to answer the question. If you don't know the answer, just say that you don't know. Use three sentences maximum and keep the answer concise."), we introduced a variation incorporating positive reinforcement \cite{Chase2022}. This modified prompt, which included elements of motivation and reward, was inspired by studies indicating that emotional stimuli and tip-offering can enhance the quality of LLM-generated responses \cite{Salinas2024, Li2023}. The revised prompt was: “You are the best assistant for question-answering tasks. Your role is to answer the question excellently using the provided context. Use the following pieces of the retrieved context to answer the question. If you don't know the answer, just say that you don't know. Use three sentences maximum and keep the answer concise. I will tip you 1000 dollars for a perfect response.” This approach resulted in a slight but measurable improvement in response quality, suggesting that motivational prompts have the potential to enhance LLM performance.

\subsection{Evaluation Using RAGAS}

To assess the performance of our enhanced RAG application, we employed the RAGAS (Retrieval-Augmented Generation Assessment System) framework, which evaluates the quality of generated outputs based on three key metrics: Context Relevance, Faithfulness, and Answer Relevance \cite{Es2023}. Each of these metrics plays a critical role in ensuring that the generated responses are accurate, relevant, and grounded in the retrieved context. The RAGAS framework utilizes the OpenAI API to automatically determine metric values, such as identifying which sentences in the retrieved context are relevant to answering the given question or which claims in the generated answer can be inferred from the given context, based on predefined prompts outlined in the RAGAS paper. Since, unfortunately, there is no benchmark test set or established ground truth available for this kind of study, we had to create a custom test set and use the RAGAS framework, which relies on LLMs in the background as evaluators. Given these limitations, this approach was the only viable option to systematically assess the performance of our proposed architecture.

Context Relevance measures the relevance of the retrieved context, which is essential because the context should contain only the information needed to answer the provided query. The metric ranges from 0 to 1, with higher values indicating better relevance. To compute Context Relevance, we identified the sentences within the retrieved context that are pertinent to the query:

\begin{equation}
\text{Context Relevance} = \frac{|S|}{\text{Total number of sentences in retrieved context}} \tag{1}
\end{equation}

Where $|S|$ is the number of sentences in the retrieved context that are relevant to answering the given question.

A high Context Relevance score indicates that the \textit{retrieved context} is highly focused and contains minimal extraneous information. This metric is particularly important in our application because our enhancements aim to optimize the retrieval and presentation of contextually accurate information. 

Faithfulness assesses the factual consistency of the textit{generated answer} with the \textit{retrieved context}. It is calculated by comparing the claims made in the generated answer with the information provided in the context. The metric also ranges from 0 to 1, with higher scores indicating better factual accuracy. This metric ensures that the generated response is grounded in the retrieved context, avoiding hallucinations and maintaining factual accuracy. For our application, Faithfulness is also crucial because it directly impacts the reliability of the responses provided to data scientists.

\begin{equation}
\text{Faithfulness score} = 
\frac{
\begin{aligned}
\text{Number of claims in the generated answer} \\
\text{that can be inferred from the given context}
\end{aligned}
}{
\text{Total number of claims in the generated answer}
}
\tag{2}
\end{equation}

Answer Relevance evaluates how well the \textit{generated answer} addresses the \textit{provided query}. This metric is assessed by computing the cosine similarity between the embeddings of the original question and the generated question, which is reverse-engineered from the answer. A higher Answer Relevance score indicates that the generated answer closely aligns with the original question, ensuring that the response is relevant and directly addresses the user's query. In our context, Answer Relevance is vital for assessing the overall effectiveness of the RAG application in producing useful and accurate responses.

\begin{equation}
\text{Answer Relevance} = \frac{1}{N} \sum_{i=1}^{N} \cos(E_{g_i}, E_o)
\tag{3}
\end{equation}

\text{Where:}
\begin{itemize}
    \item \( E_{g_i} \) is the embedding of the generated question \( i \).
    \item \( E_o \) is the embedding of the original question.
    \item \( N \) is the number of generated questions
\end{itemize}

\begin{figure}[ht]
    \centering
    \includegraphics[width=0.6\linewidth]{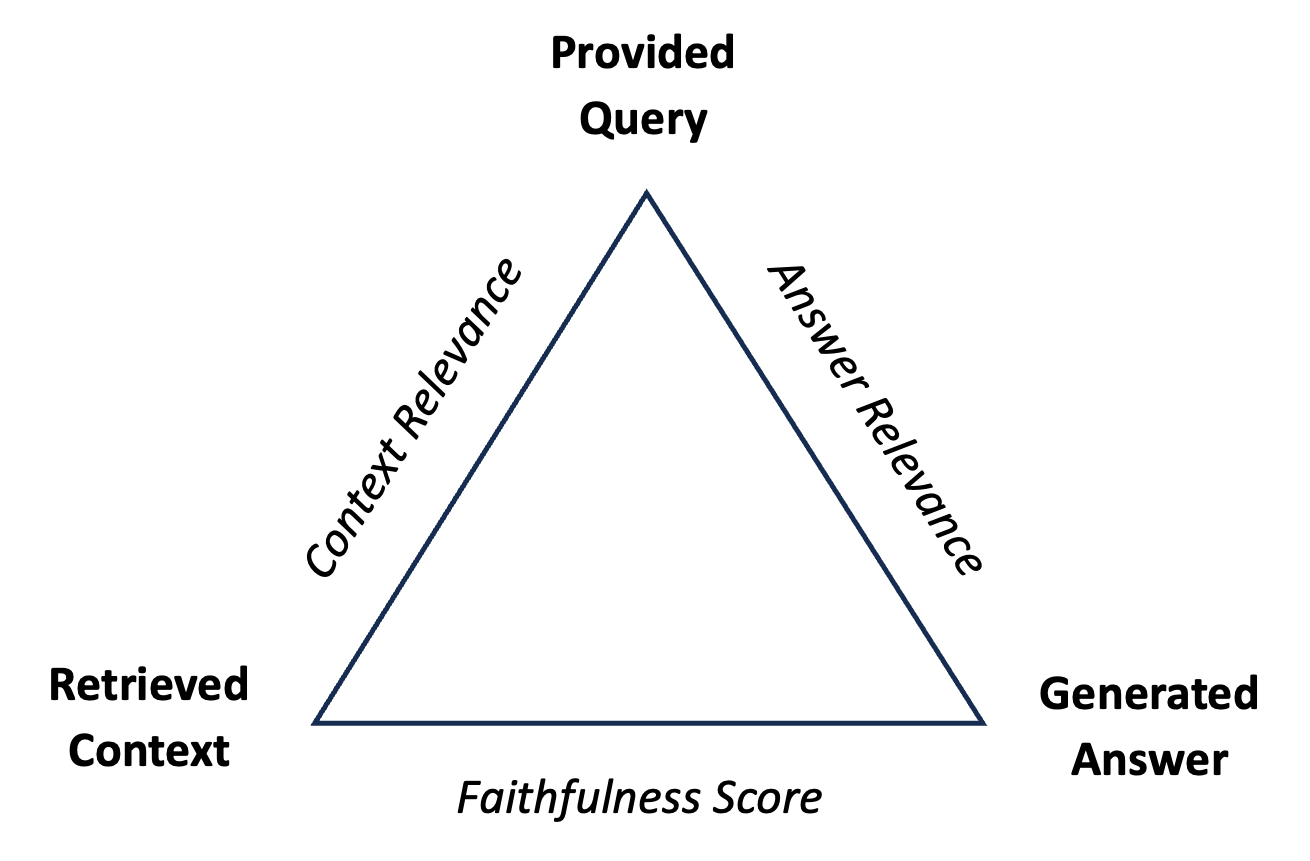}
    \captionsetup{font=small} % Set font size for this caption
    \caption{Three Performance Metrics of RAGAS Framework and Their
Corresponding Aspects}
    \label{fig:Metrics}
\end{figure}

Using these detailed metrics, the RAGAS framework provides a comprehensive evaluation of the performance of our enhanced RAG system. Each metric measures performance in different but complementary areas. Figure \ref{fig:Metrics} visually illustrates the aspects that these metrics cover. The metrics not only validate the improvements we have made but also offer insights into areas where further enhancements can be pursued, ensuring that the application continues to deliver high-quality, reliable, and contextually relevant answers to data scientists.

We have also incorporated the average word count into the analysis and displayed it in the results table to consider its impact on these metrics. Specifically, metrics like context relevance are highly dependent on the number of words in the retrieved contexts. When more sentences are included in a context, it becomes more challenging to achieve high scores across these metrics because, for context relevance to be maximized, all sentences should be relevant to the question. With more sentences, the likelihood of including unrelated content increases, potentially lowering the overall score.

\section{Results}

The results of our experiments, summarized in the tables, demonstrate the impact of various enhancements to the RAG pipeline on key metrics: Context Relevance (CR), Faithfulness, and Answer Relevance. Each metric's value is averaged over 1,500 replications -30 replications for each one of the 50 questions used in the analysis. The choice of 30 replications per question is grounded in the Central Limit Theorem (CLT), which asserts that with a sample size of 30, the distribution of the sample means will approximate a normal distribution, ensuring the robustness of our results  \cite{Moore1989}.

\subsection{Performance metrics for different fine-tuning experiments}

\begin{table}[H]
\centering
\captionsetup{font=small} % Set the font size for the caption
\caption{Performance metrics for different fine-tuning experiments. (FT = Fine-tuning, TB = Textbooks, G = GROBID, SNS = Semantic Node Splitter)}
\label{tab:results1}
\small % Adjust font size
\begin{tabularx}{\textwidth}{|l|X|X|X|X|}
\hline
\textbf{Experiment} & \textbf{Context} \newline \textbf{Relevancy} & \textbf{Faithfulness} & \textbf{Answer} \newline \textbf{Relevance} & \textbf{Average Word Count} \\ \hline
1. No FT & 0.302 & 0.687 & 0.888 & 853.88 \\ \hline
2. 5 TB & 0.337 & 0.658 & 0.871 & 853.90 \\ \hline
3. 17 TB & {\textbf{0.372 }}& {\textbf{0.756}} & 0.902 & 836.10 \\ \hline
4. 5 TB + G & 0.062 & 0.198 & 0.790 & 913.00 \\ \hline
5. 17 TB + G & 0.335 & 0.740 & {\textbf{0.909}} & 839.34 \\ \hline
6. 17 TB + G + SNS & 0.339 & 0.701 & 0.902 & 841.72 \\ \hline
\end{tabularx}
\end{table}

In Table \ref{tab:results1}, all experiments, including \textit{No FT} case which corresponds to \textit{B + G} in Table \ref{tab:results2}), the models incorporated GROBID for preprocessing and fine-tuning to standardize data structuring. Variations in performance were solely driven by differences in the fine-tuning data, particularly the amount and type of training materials. Without fine-tuning, the baseline model incorporated with GROBID exhibited relatively low Context Relevance (CR) at 0.0302, despite relatively high Faithfulness (0.687) and Answer Relevance (0.888). This highlighted the necessity of fine-tuning to improve the model's ability to retrieve relevant content.

Fine-tuning on 5 textbooks (Experiment 2) significantly improved CR to 0.337, with a slight rise in Faithfulness and a small decline in Answer Relevance. When expanded to 17 textbooks (Experiment 3), CR reached 0.372, demonstrating the advantage of a larger training dataset, with improved Faithfulness (0.756) and Answer Relevance (0.902). This model was selected for further testing due to its strong overall performance in CR.

In Experiment 4, adding GROBID to fine-tune with 5 textbooks caused a drop in CR to 0.062, likely due to the removal of important content during preprocessing. However, combining GROBID with fine-tuning with 17 textbooks (Experiment 5) restored CR to 0.335, improving both Faithfulness and Answer Relevance. Further improvements were seen in Experiment 6, where the addition of Semantic Node Splitter (SNS) slightly increased CR to 0.339, maintaining competitive Faithfulness and high Answer Relevance.

Overall, fine-tuning with larger datasets significantly improved CR, while GROBID and SNS added further refinements. The models from Experiments 3 (17 TB) and 6 (17 TB + G + SNS) were chosen for further evaluation, having demonstrated the best performance in retrieving relevant academic context, which is the main focus of this study.

\begin{table}[H]
\centering
\captionsetup{font=small}
\caption{Significant Pairwise Comparisons for Context Relevance using Tukey's HSD Test.}
\label{tab:significance_cr}
\small
\begin{tabularx}{\textwidth}{|l|X|X|X|}
\hline
\textbf{Comparison} & \textbf{Mean} \newline \textbf{Difference} & \textbf{p-value} & \textbf{Significant?} \\ \hline
1 vs. 2 & 0.035 & $<$0.001 & Yes \\ \hline
2 vs. 3 & 0.035 & $<$0.001 & Yes \\ \hline
3 vs. 4 & -0.310 & $<$0.001 & Yes \\ \hline
3 vs. 5 & -0.037 & $<$0.001 & Yes \\ \hline
3 vs. 6 & -0.034 & $<$0.001 & Yes \\ \hline
6 vs. 2 & -0.002 & 0.999 & No \\ \hline
6 vs. 5 & -0.003 & 0.994 & No \\ \hline
2 vs. 5 & -0.002 & 0.9996 & No \\ \hline
\end{tabularx}
\end{table}

Context Relevance was utilized exclusively to perform the significance tests because the primary aim is to assess the impact of fine-tuning on the model's ability to retrieve relevant content and adapt the embedding model effectively. Since it is crucial for the embedding to accurately understand the query and obtain the relevant chunks, focusing on Context Relevance allows us to detect the contributions of different fine-tuning experiments more precisely. To statistically evaluate the significance of the differences in Context Relevance across configurations, we employed Welch's ANOVA followed by Tukey's HSD test for post-hoc analysis \cite{Welch1951} \cite{Tukey1949}. Welch's ANOVA was chosen because it does not assume equal variances, making it a more robust method when the homogeneity of variances assumption is violated. This approach ensures the reliability of the results by accounting for variance differences across the experiments. Tukey's HSD test was then applied to identify which specific pairwise comparisons showed statistically significant differences, providing a detailed understanding of the performance variations between the configurations.

The results of Tukey's HSD tests presented in Table \ref{tab:significance_cr} indicate a statistically significant (with p$<$0.001) enhancement in Context Relevance from the baseline Experiment 1 (No FT) to Experiment 3 (17 TB). Evidently, Experiment 3 exhibits a statistically significant superiority over all other experimental conditions concerning Context Relevance. While the results for Experiment 6 (17 TB + G + SNS), Experiment 2 (5 TB), and Experiment 5 (17 TB + G) are closely aligned, exhibiting no statistically significant differences among them as detailed in the last three rows of Table \ref{tab:significance_cr}, we have chosen to further evaluate Experiment 6 alongside Experiment 3 due to its position as the second highest in overall average Context Relevance.

\subsection{Performance metrics for RAG pipeline configurations using the fine-tuning model with Experiment 3}

\begin{table}[ht]
\centering
\captionsetup{font=small} % Set the font size for the caption
\caption{Performance metrics for different RAG pipeline configurations using the fine-tuning model with Experiment 3. (B = Baseline, G = GROBID, FT = Fine-tuning, SC = Semantic Chunking, AF = Abstract First, EPT = Enhanced Prompting Technique)}
\label{tab:results2}
\small % Adjust font size
\begin{tabularx}{\textwidth}{|l|X|X|X|X|}
\hline
\textbf{Configuration} & \textbf{Context} \newline \textbf{Relevancy} & \textbf{Faithfulness} & \textbf{Answer} \newline \textbf{Relevance} & \textbf{Average Word Count} \\ \hline
1. B & 0.031 & 0.622 & 0.898 & 772.92 \\ \hline
2. B + G & 0.302 & 0.687 & 0.888 & 853.88 \\ \hline
3. B + G + FT & 0.372 & {\textbf{0.756}} & {\textbf{0.902}} & 836.10 \\ \hline
4. B + G + FT + SC & {\textbf{0.456 }}& 0.599 & 0.852 & 947.94 \\ \hline
5. B + G + FT + SC + AF & 0.447 & 0.606 & 0.848 & 915.80 \\ \hline
6. B + G + FT + SC + AF + EPT & 0.454 & 0.588 & 0.887 & 917.86 \\ \hline
\end{tabularx}
\end{table}

The performance metrics for different RAG pipeline configurations using the fine-tuned model from Experiment 3, as shown in Table \ref{tab:results2}, illustrate how various enhancements affect the retrieval process. Adding GROBID (B + G) significantly improved Context Relevance (CR) from 0.031 to 0.302, demonstrating its effectiveness in structuring the data. Faithfulness increased to 0.687, while Answer Relevance slightly declined to 0.888, indicating some variability in the use of the retrieved context. The average word count also rose to 853.88, reflecting broader context retrieval with GROBID.

Combining GROBID with fine-tuning (B + G + FT) further improved CR to 0.372, with higher Faithfulness (0.756) and Answer Relevance (0.902). This configuration had the highest Faithfulness, suggesting efficient use of the relevant context. The average word count dropped to 836.10, reflecting a more refined retrieval process.

Semantic Chunking (B + G + FT + SC) raised CR to 0.456, improving retrieval of semantically coherent chunks, though Faithfulness dropped to 0.599, likely due to the larger average word count (947.94). Adding the Abstract-First strategy (B + G + FT + SC + AF) slightly reduced CR to 0.447 but improved Faithfulness to 0.606, with a lower average word count of 915.80, indicating more concise retrieval.

The final configuration (B + G + FT + SC + AF + EPT) maintained a high CR (0.454) and improved Answer Relevance to 0.887. Faithfulness slightly decreased to 0.588, suggesting the potential for better alignment between context and generated responses. Overall, the fine-tuning model from Experiment 2 performed best when GROBID, fine-tuning, and Semantic Chunking were combined, while Enhanced Prompting improved the quality of responses.

\begin{table}[ht]
\centering
\captionsetup{font=small}
\caption{Significant pairwise comparisons for different metrics using Tukey's HSD test for Table 3. (CR = Context Relevance, F = Faithfulness, AR = Answer Relevance)}
\label{tab:significance_all_metrics_exp3}
\small
\begin{tabularx}{\textwidth}{|l|l|X|X|X|}
\hline
\textbf{Metric} & \textbf{Comparison} & \textbf{Mean} \newline \textbf{Difference} & \textbf{p-value} & \textbf{Significant?} \\ \hline
\multirow{5}{*}{\textbf{Context Relevance}} 
    & 1 vs. 2 & 0.2713 & $<$0.001 & Yes \\ \cline{2-5}
    & 2 vs. 3 & 0.0704 & $<$0.001 & Yes \\ \cline{2-5}
    & 3 vs. 4 & 0.0838 & $<$0.001 & Yes \\ \cline{2-5}
    & 4 vs. 5 & -0.0089 & 0.952 & No \\ \cline{2-5}
    & 5 vs. 6 & 0.0073 & 0.980 & No \\ \hline
\multirow{5}{*}{\textbf{Faithfulness}} 
    & 1 vs. 2 & 0.0651 & $<$0.001 & Yes \\ \cline{2-5}
    & 2 vs. 3 & 0.0691 & $<$0.001 & Yes \\ \cline{2-5}
    & 3 vs. 4 & -0.1576 & $<$0.001 & Yes \\ \cline{2-5}
    & 4 vs. 5 & 0.0068 & 0.992 & No \\ \cline{2-5}
    & 5 vs. 6 & -0.0171 & 0.673 & No \\ \hline
\multirow{5}{*}{\textbf{Answer Relevance}} 
    & 1 vs. 2 & -0.0103 & 0.454 & No \\ \cline{2-5}
    & 2 vs. 3 & 0.0141 & 0.126 & No \\ \cline{2-5}
    & 3 vs. 4 & -0.0496 & $<$0.001 & Yes \\ \cline{2-5}
    & 4 vs. 5 & -0.0036 & 0.989 & No \\ \cline{2-5}
    & 5 vs. 6 & 0.0380 & $<$0.001 & Yes \\ \hline
\end{tabularx}
\end{table}

\FloatBarrier 

The significant pairwise comparisons presented in Table \ref{tab:significance_all_metrics_exp3} illustrate the impact of each RAG pipeline enhancement on the key performance metrics. Until the fourth configuration, the successive improvements in Context Relevance and Faithfulness were statistically significant, as evidenced by the p-values below 0.001 for each step. This indicates that adding GROBID, fine-tuning, and Semantic Chunking sequentially contributed meaningfully to the improvement in these metrics. However, the Abstract-First strategy and Enhanced Prompting Technique (configurations 5 and 6) did not produce statistically significant changes in Context Relevance or Faithfulness, suggesting that these strategies may not further optimize these aspects. In contrast, Answer Relevance only showed significant differences in configurations 4 and 6, corresponding to the application of Semantic Chunking and the Enhanced Prompting Technique. This is consistent with the role of Semantic Chunking in retrieving more relevant context, thereby improving the relevance of the generated responses, and the Enhanced Prompting Technique’s ability to refine the model's output, yielding more precise answers to the posed questions.

\subsection{Performance metrics for RAG pipeline configurations using the fine-tuning model with Experiment 6}

\begin{table}[H]
\centering
\captionsetup{font=small} % Set the font size for the caption
\caption{Performance metrics for different RAG pipeline configurations using the fine-tuning model with Experiment 6. (B = Baseline, G = GROBID, FT = Fine-tuning, SC = Semantic Chunking, AF = Abstract First, EPT = Enhanced Prompting Technique)}
\label{tab:results3}
\small % Adjust font size
\begin{tabularx}{\textwidth}{|l|X|X|X|X|}
\hline
\textbf{Configuration} & \textbf{Context} \newline \textbf{Relevancy} & \textbf{Faithfulness} & \textbf{Answer} \newline \textbf{Relevance} & \textbf{Average Word Count} \\ \hline
1. B & 0.031 & 0.622 & 0.898 & 772.92 \\ \hline
2. B + G & 0.302 & 0.687 & 0.888 & 853.88 \\ \hline
3. B + G + FT & 0.339 & {\textbf{0.701}} & {\textbf{0.902}} & 841.72 \\ \hline
4. B + G + FT + SC & 0.475 & 0.618 & 0.872 & 840.24 \\ \hline
5. B + G + FT + SC + AF & {\textbf{0.507}} & 0.592 & 0.863 & 753.16 \\ \hline
6. B + G + FT + SC + AF + EPT & {\textbf{0.507}} & 0.592 & 0.890 & 753.16 \\ \hline
\end{tabularx}
\end{table}

%% \clearpage
 
The results in Table \ref{tab:results3} present the performance of various RAG pipeline configurations using the fine-tuned model from Experiment 6 across key metrics. The baseline (B) configuration shows low Context Relevance (CR) at 0.031, with higher Faithfulness (0.622) and Answer Relevance (0.898), but the retrieved context remains minimally effective. Adding GROBID (B + G) improves CR to 0.302, with a slight increase in Faithfulness (0.687) and a small decrease in Answer Relevance (0.888). The average word count also rises to 853.88, reflecting broader content retrieval.

Introducing fine-tuning alongside GROBID (B + G + FT) further increases CR to 0.339, with Faithfulness rising to 0.701 and Answer Relevance stabilizing at 0.902. Semantic Chunking (B + G + FT + SC) brings the highest CR yet (0.485), though Faithfulness drops slightly to 0.640, with Answer Relevance decreasing to 0.874. The Abstract-First strategy (B + G + FT + SC + AF) improves CR to 0.507, though both Faithfulness (0.592) and Answer Relevance (0.863) decrease, possibly due to the narrower focus on abstract-level content.

Finally, integrating the Enhanced Prompting Technique (B + G + FT + SC + AF + EPT) maintains CR at 0.507 while increasing Answer Relevance to 0.890. This suggests that enhanced prompts help the model generate more coherent responses, even as Faithfulness remains stable at 0.592. Overall, the best performance in CR was achieved with Semantic Chunking and Abstract-First strategies, while enhanced prompting improved answer quality.

\begin{table}[H]
\centering
\captionsetup{font=small}
\caption{Significant pairwise comparisons for different metrics using Tukey's HSD test for Table 5. (CR = Context Relevance, F = Faithfulness, AR = Answer Relevance)}
\label{tab:significance_all_metrics_exp6}
\small
\begin{tabularx}{\textwidth}{|l|l|X|X|X|}
\hline
\textbf{Metric} & \textbf{Comparison} & \textbf{Mean} \newline \textbf{Difference} & \textbf{p-value} & \textbf{Significant?} \\ \hline
\multirow{5}{*}{\textbf{Context Relevance}} 
    & 1 vs. 2 & 0.2713 & $<$0.001 & Yes \\ \cline{2-5}
    & 2 vs. 3 & 0.0367 & 0.0065 & Yes \\ \cline{2-5}
    & 3 vs. 4 & 0.1374 & $<$0.001 & Yes \\ \cline{2-5}
    & 4 vs. 5 & 0.0315 & 0.0342 & Yes \\ \cline{2-5}
    & 5 vs. 6 & -0.0002 & 1.0000 & No \\ \hline
\multirow{5}{*}{\textbf{Faithfulness}} 
    & 1 vs. 2 & 0.0651 & $<$0.001 & Yes \\ \cline{2-5}
    & 2 vs. 3 & 0.0137 & 0.834 & No \\ \cline{2-5}
    & 3 vs. 4 & -0.0868 & $<$0.001 & Yes \\ \cline{2-5}
    & 4 vs. 5 & -0.0224 & 0.3643 & No \\ \cline{2-5}
    & 5 vs. 6 & 0.0002 & 1.0000 & No \\ \hline
\multirow{5}{*}{\textbf{Answer Relevance}} 
    & 1 vs. 2 & -0.0103 & 0.3491 & No \\ \cline{2-5}
    & 2 vs. 3 & 0.0140 & 0.0736 & No \\ \cline{2-5}
    & 3 vs. 4 & -0.0295 & $<$0.001 & Yes \\ \cline{2-5}
    & 4 vs. 5 & -0.0090 & 0.5050 & No \\ \cline{2-5}
    & 5 vs. 6 & 0.0270 & $<$0.001 & Yes \\ \hline
\end{tabularx}
\end{table}

The significant pairwise comparisons in Table \ref{tab:significance_all_metrics_exp6} indicate consistent improvements in Context Relevance across each step, up until the final configuration involving Enhanced Prompting. This aligns with expectations, as Enhanced Prompting primarily influences the response generation phase rather than context retrieval. The Abstract-First approach (configuration 5) produced a significant enhancement in Context Relevance, contrasting with its lack of significant effect in Experiment 3 (Table \ref{tab:significance_all_metrics_exp3}). This suggests that Abstract-First may facilitate more efficient retrieval processes in certain experimental conditions. Faithfulness significantly increased only with the initial addition of GROBID (configuration 2), while it showed a notable decline with the introduction of Semantic Chunking (configuration 4), possibly due to the increased length of retrieved content impacting alignment with the generated response. For Answer Relevance, the metric significantly decreased with the addition of Semantic Chunking but later improved with Enhanced Prompting, reflecting the latter's role in refining the generated responses to align more closely with the retrieved context.

\section{Discussion}

The experiments conducted in this study aimed to refine the Retrieval-Augmented Generation (RAG) pipeline for academic literature, focusing on enhancing the retrieval of context relevant to data science-related queries. Several configurations were evaluated across key metrics, including Context Relevance (CR), Faithfulness, and Answer Relevance, providing insight into how different components of the pipeline affect its overall performance. The emphasis of this work was on improving CR, given its direct correlation with the ability to retrieve pertinent context from large corpora of academic literature.

One of the key findings is the positive impact of fine-tuning on CR across all configurations. Models that were fine-tuned using domain-specific textbooks, particularly in Experiments 2 and 5, consistently outperformed the baseline model in retrieving context relevant to the query. This highlights the importance of curating high-quality, domain-specific datasets when fine-tuning language models to optimize their performance for specialized tasks. The results also suggest that increasing the size of the fine-tuning corpus, as seen in Experiment 2, improves not only CR but also Faithfulness and Answer Relevance. This trend underscores the value of larger and more comprehensive training datasets for improving the system's ability to retrieve and use relevant information.

The introduction of GROBID provided mixed results. While it significantly improved CR by structuring the data for more precise retrieval, its incorporation without sufficient fine-tuning, as seen in Experiment 3, led to a noticeable drop in CR. This suggests that while GROBID's document structuring is beneficial, it must be paired with ample training data to ensure that critical context is not inadvertently removed during preprocessing. The combination of GROBID with fine-tuning on larger datasets restored CR to higher levels, as observed in Experiments 4 and 5, indicating that the issues introduced by GROBID can be mitigated with appropriate data preparation \cite{Lopez2024}.

Semantic Chunking also played a significant role in improving CR, particularly in the final stages of the pipeline. By grouping text into semantically coherent chunks, this technique ensured that the model retrieved more meaningful and contextually relevant information. However, the trade-off between longer retrieved contexts and Faithfulness suggests that future work could explore ways to balance chunk size with the system's ability to generate precise answers from the retrieved information. This is evident in the slight reduction in Faithfulness observed when Semantic Chunking was introduced.

The Abstract-First strategy introduced in the pipeline further refined the retrieval process by focusing on abstracts to filter relevant content before conducting a more in-depth full-text search. This approach led to notable improvements in CR, achieving the highest scores across all configurations. Interestingly, the use of abstracts challenges the conventional assumption that broader searches yield better results. The findings suggest that a more focused search using abstract-level information can be equally, if not more, effective than full-text searches. This method not only reduced the average word count but also demonstrated that a concise, abstract-based search can still maintain high relevance in context retrieval. However, it remains to be seen whether this strategy is as effective in other domains where abstracts may not fully capture the nuances of the article’s content, presenting an area for future research.

The Enhanced Prompting Technique, while not affecting CR directly, improved Answer Relevance by guiding the LLM to better utilize the retrieved context when generating answers. Since the LLM combines the retrieved context with the query to produce the response, the prompt’s structure plays a pivotal role in maximizing the utility of the language model. The relatively stable CR across configurations that employed enhanced prompting confirms that this technique’s impact is not on retrieval but rather on improving how well the model aligns its responses with the given context. This result underscores the importance of prompt design in optimizing response generation. Moreover, there remains room for further improvements in prompt alignment, as the open-ended nature of prompts offers opportunities to refine how context and query are combined to generate more precise and contextually accurate responses.

In future enhancements, incorporating knowledge graph-based approaches for automating the validation and structuring of retrieved content could significantly improve the accuracy and reliability of RAG applications. Such techniques, as demonstrated in the construction and validation of behavioral models using automated knowledge extraction, show promise for dynamically synthesizing robust and relevant information from diverse sources \cite{Sonnenschein2024}. By integrating these methods, RAG systems could better manage complex information retrieval tasks, thereby further enhancing decision-making processes across various domains.

\section{Conclusion}

In conclusion, this study illustrates that the proposed Enhanced RAG Architecture, which integrates fine-tuning on domain-specific datasets, employs GROBID for data structuring, implements Semantic Chunking, and utilizes an Abstract-First strategy, significantly enhances the capability of the RAG pipeline to retrieve pertinent academic content. The fine-tuning models with larger and more diverse training sets, as demonstrated in Experiments 3 and 6, were particularly effective in increasing Context Relevance (CR), which aligns with the study's goal of improving retrieval of academic literature for data scientists. While Semantic Chunking and Abstract-First approaches further refined retrieval precision, the observed trade-offs in Faithfulness suggest a need for future work to optimize how retrieved content is utilized in generating responses. The Enhanced Prompting Technique contributed to more contextually aligned answers, highlighting the role of prompt design in enhancing LLM performance within RAG systems. These findings offer useful insights for advancing context retrieval systems in academic research, with further investigations needed to assess the broader applicability of these methods.

Despite its contributions, this study has several limitations.
One key challenge is the lack of a standardized ground truth for evaluating the quality of the generated answers. To address this, we utilized the RAGAS framework, which relies on LLMs for evaluation. While this method was the most viable given the circumstances, it introduces a degree of subjectivity into the assessment, potentially affecting the interpretation of the results. Additionally, the reliance on a custom test set, developed specifically for this study, poses another limitation. Although the questions were designed to reflect realistic challenges in data science, the lack of a widely accepted benchmark for RAG evaluation in academic literature restricts the generalizability of the findings. Future research should aim to establish standardized benchmarks and ground truth datasets to facilitate more robust and comparable evaluations across studies.

%\section*{Acknowledgments}
\label{sec:sample:appendix}

% Acknowledgments text here.

\appendix

\section*{Appendix A. Advanced Data Science Questions}
\label{AppendixA}

\begin{enumerate}
    \item I'm a data scientist in the financial sector, selecting features to predict stock prices. I've used traditional methods like backward elimination. Which advanced feature selection technique should I consider to enhance model performance?
    \item In the context of agriculture, I'm predicting crop yields based on soil and weather data. Having tried manual feature selection, what automated feature selection method would help in identifying the most influential factors for yield prediction?
    \item As a data scientist in the insurance industry, I'm working on customer churn prediction. I've used correlation-based feature selection but need to account for nonlinear relationships. Which technique would be suitable for this purpose?
    \item Working in the healthcare sector, I deal with patient vital signs data that often contains outliers. After using the Z-score for outlier detection, what robust outlier elimination technique should I adopt for cleaner datasets?
    \item In the manufacturing industry, I'm analyzing sensor data from machinery for predictive maintenance. Having used the IQR method to detect outliers, which advanced technique would you recommend for more accurate anomaly detection and elimination?
    \item I'm analyzing customer behavior data in e-commerce, using PCA for dimensionality reduction. Can you suggest a modern technique that might offer better insights while preserving data variance?
    \item In environmental science, I deal with high-dimensional climate data. Having used t-SNE for visualization, what other dimensionality reduction method should I consider for more effective data analysis?
    \item As a data scientist in the telecom sector, I work with extensive network traffic data. After applying LDA for feature extraction, what newer technique would enhance the interpretability and accuracy of my models?
    \item In the energy industry, I'm analyzing sensor data from smart grids. While PCA has helped in reducing dimensions, what alternative should I explore for better performance with large datasets?
    \item I'm studying brain imaging data for neurological research. Having tried factor analysis, what other dimensionality reduction technique could help in uncovering hidden patterns more effectively?
    \item In the automotive industry, I develop models for autonomous vehicles using real-world driving data. What synthetic data generation technique can I use to augment my dataset while ensuring realistic scenarios?
    \item Working with confidential financial data, I need to generate synthetic datasets for model training. Which method would provide a balance between data utility and privacy?
    \item As a data scientist in social sciences, I deal with sensitive survey data. What technique should I use to generate synthetic data that maintains the statistical properties of the original data?
    \item In robotics, I'm training a robot to navigate complex environments. Having used Q-learning, what advanced reinforcement learning algorithm should I explore for improved performance in dynamic settings?
    \item As a game developer, I'm using reinforcement learning to create intelligent agents. Beyond DQN, which algorithm would enhance the adaptability and learning speed of my agents?
    \item In the pharmaceutical industry, I'm classifying drug responses based on genetic data. After using SVM, what advanced classification algorithm should I consider for better accuracy and interpretability?
    \item As a data scientist in the automotive sector, I'm classifying vehicle types from sensor data. Having used decision trees, which more sophisticated method could improve classification performance?
    \item Working in cybersecurity, I classify network traffic for threat detection. Beyond random forests, what advanced classification technique should I employ for more accurate threat identification?
    \item In the education sector, I'm classifying student performance based on learning behaviors. After using k-NN, which algorithm would offer better results considering interpretability and accuracy?
    \item As a data scientist in the legal domain, I'm classifying legal documents by type. What advanced classification method would enhance my document categorization system?
    \item In the energy sector, I'm predicting energy consumption using multiple regression. What advanced regression model could provide better accuracy and handle nonlinearity more effectively?
    \item Working in the telecommunications industry, I'm predicting customer call durations. Having tried ridge regression, which sophisticated regression technique should I explore next?
    \item In the field of epidemiology, I'm predicting disease spread based on environmental factors. Beyond polynomial regression, what model should I use to improve predictive accuracy?
    \item As a data scientist in real estate, I'm predicting property values. After using lasso regression, which advanced method would you recommend for incorporating spatial dependencies?
    \item In transportation, I'm predicting traffic flow using regression models. What advanced technique should I consider for capturing complex patterns in traffic data?
    \item In the context of smart cities, I'm clustering IoT sensor data. Having used DBSCAN, what other clustering algorithm would be effective for large, diverse datasets?
    \item As a data scientist in the food industry, I'm clustering consumer taste preferences. Beyond hierarchical clustering, which algorithm should I explore for better segmentation?
    \item Working in the entertainment sector, I'm clustering movie preferences. What advanced clustering method can help in identifying more nuanced audience segments?
    \item In environmental monitoring, I'm clustering air quality data. After using agglomerative clustering, which technique would provide better results for large-scale environmental datasets?
    \item As a data scientist in sports analytics, I'm clustering player performance metrics. Beyond Gaussian Mixture Models, which advanced clustering method should I consider for improved insights?
    \item In medical imaging, I'm detecting abnormalities in MRI scans. Beyond CNNs, what advanced deep learning architecture should I explore for higher accuracy?
    \item As a data scientist in agriculture, I'm analyzing drone imagery for crop health. Having used image segmentation, which technique should I adopt for a more precise analysis?
    \item Working with satellite imagery in environmental science, I'm classifying land use. Beyond traditional CNNs, which model should I explore for improved accuracy and efficiency?
    \item In the automotive sector, I'm developing object detection systems for autonomous vehicles. What advanced image processing technique would enhance detection accuracy in real-time scenarios?
    \item As a data scientist in fashion, I'm analyzing images for style recognition. Beyond transfer learning, which approach should I use to improve model performance?
    \item In legal tech, I'm analyzing contracts using NLP. Having used word embeddings, what advanced technique should I consider for better understanding legal jargon and nuances?
    \item As a data scientist in customer service, I'm developing chatbots for automated responses. Beyond basic sequence models, which advanced NLP model should I explore for more natural interactions?
    \item Working in publishing, I'm analyzing book reviews for sentiment analysis. What state-of-the-art NLP technique would help in capturing complex sentiments more accurately?
    \item In the field of education, I'm developing automated grading systems. Having used traditional NLP methods, which advanced model should I adopt for better accuracy in grading essays?
    \item As a data scientist in social media analysis, I'm detecting trends from tweets. Beyond LSTM, which cutting-edge NLP technique should I consider for more insightful trend analysis?
    \item In finance, I'm forecasting stock prices using traditional ARIMA models. What advanced time series analysis method should I use to better capture market volatility?
    \item As a data scientist in retail, I'm predicting sales trends. Beyond SARIMA, which technique should I explore to account for seasonal variations and promotions?
    \item Working in the energy sector, I'm forecasting electricity demand. Having used exponential smoothing, what advanced method should I consider for more accurate demand predictions?
    \item In healthcare, I'm analyzing patient vital signs over time. What advanced time series model would enhance the predictive accuracy of health outcomes?
    \item As a data scientist in transportation, I'm forecasting passenger flow. Beyond traditional methods, which advanced technique should I adopt to improve forecasting accuracy?
    \item In the context of fraud detection, I'm evaluating the performance of my models. Beyond accuracy and precision, what other evaluation metrics should I consider for a more comprehensive assessment?
    \item As a data scientist in healthcare, I'm evaluating models for disease prediction. Which evaluation method would help in understanding the trade-offs between sensitivity and specificity more effectively?
    \item Working in marketing, I'm assessing the effectiveness of customer segmentation models. Beyond the silhouette score, what other evaluation techniques should I use to validate the quality of my clusters?
    \item In the field of education, I'm evaluating models for predicting student success. What advanced evaluation metrics should I consider to ensure my models are both accurate and fair?
    \item As a data scientist in manufacturing, I'm assessing the performance of predictive maintenance models. Beyond traditional metrics, which evaluation method should I use to measure the real-world impact of my models?
\end{enumerate}

\section*{Appendix B. 5 Textbooks}
\label{AppendixB}

\begin{enumerate}
    \item Aggarwal, C. C. (2018). \textit{Neural Networks and Deep Learning: A Textbook}. Springer.
    \item Alpaydın, E. (2014). \textit{Introduction to Machine Learning} (3rd ed.). The MIT Press.
    \item Bruce, P., \& Bruce, A. (2017). \textit{Practical Statistics for Data Scientists: 50 Essential Concepts}. O'Reilly Media.
    \item Langr, J., \& Bok, V. (2019). \textit{GANs in Action: Deep Learning with Generative Adversarial Networks}. Manning Publications.
    \item Montgomery, D. C., Jennings, C. L., \& Kulahci, M. (2015). \textit{Introduction to Time Series Analysis and Forecasting} (2nd ed.). Wiley.
\end{enumerate}

\section*{Appendix C. 17 Textbooks}
\label{AppendixC}

\begin{enumerate}
    \item Aßenmacher, Matthias. \textit{Multimodal Deep Learning}. Self-published, 2023.
    \item Bertsekas, Dimitri P. \textit{A Course in Reinforcement Learning}. Arizona State University.
    \item Boykis, Vicki. \textit{What are Embeddings}. Self-published, 2023.
    \item Bruce, Peter, and Andrew Bruce. \textit{Practical Statistics for Data Scientists: 50 Essential Concepts}. O'Reilly Media, 2017.
    \item Daumé III, Hal. \textit{A Course in Machine Learning}. Self-published.
    \item Deisenroth, Marc Peter, A. Aldo Faisal, and Cheng Soon Ong. \textit{Mathematics for Machine Learning}. Cambridge University Press, 2020.
    \item Devlin, Hannah, Guo Kunin, Xiang Tian. \textit{Seeing Theory}. Self-published.
    \item Gutmann, Michael U. \textit{Pen \& Paper: Exercises in Machine Learning}. Self-published.
    \item Jung, Alexander. \textit{Machine Learning: The Basics}. Springer, 2022.
    \item Langr, Jakub, and Vladimir Bok. \textit{Deep Learning with Generative Adversarial Networks}. Manning Publications, 2019.
    \item MacKay, David J.C. \textit{Information Theory, Inference, and Learning Algorithms}. Cambridge University Press, 2003.
    \item Montgomery, Douglas C., Cheryl L. Jennings, and Murat Kulahci. \textit{Introduction to Time Series Analysis and Forecasting}. 2nd Edition, Wiley, 2015.
    \item Nilsson, Nils J. \textit{Introduction to Machine Learning: An Early Draft of a Proposed Textbook}. Stanford University, 1996.
    \item Prince, Simon J.D. \textit{Understanding Deep Learning}. Draft Edition, 2024.
    \item Shashua, Amnon. \textit{Introduction to Machine Learning}. The Hebrew University of Jerusalem, 2008.
    \item Sutton, Richard S., and Andrew G. Barto. \textit{Reinforcement Learning: An Introduction}. 2nd Edition, MIT Press, 2018.
    \item Alpaydin, Ethem. \textit{Introduction to Machine Learning}. 3rd Edition, MIT Press, 2014.
\end{enumerate}

\newpage
\section*{Appendix D. Sample Output For Baseline Model for Question 40}
\label{AppendixD}

\begin{center}
\renewcommand{\arraystretch}{1} % Adjust row height
\tiny
\begin{longtable}{|p{1cm}|p{8cm}|p{2.5cm}|p{1.3cm}|p{1.35cm}|p{1.3cm}|}
\hline
%\centering \textbf{Question} & \centering \textbf{Contexts} & \centering \textbf{Answer} & \centering \textbf{Context Relevance}  & \centering \textbf{Faithfulness} & \centering \textbf{Answer Relevance} \\\hline
\textbf{Question}&\textbf{Contexts}&\textbf{Answer}&\textbf{Context Relevance}&\textbf{Faithfulness}&\textbf{Answer Relevance}\\\hline
{\setlength{\baselineskip}{0.8\baselineskip} As a data scientist in social media analysis, I'm detecting trends from tweets. Beyond LSTM, which cutting-edge NLP technique should I consider for more insightful trend analysis?} & {\setlength{\baselineskip}{0.8\baselineskip} •ample, in the 2016 U.S elections, along with \#Election2016, the two political parties used the \#VoteTrump and \#ivotedforHillaryClinton respectively to reach out to the general public on social media. In Twitter, for a hashtag to qualify as a trending topic, the Twitter trending algorithm is designed in such a way that the users will be shown the trending tweets based on their location, interests and the profiles they follow. Trends are usually the hashtags that have been used vastly in a short period. Facebook also has a trending module to show what topics are being discussed. Trends are an adequate way to ignite a discussion on a topic. However, considering the dynamic and erratic nature of social media, a trending topic can be replaced within seconds of its initiation; this can have a disadvantage over a current problem, which is trending. If the issue at hand carries a significant concern and needs to reach out to its con-

•The engagement for Vauxhall remained at a consistent
level throughout the event (see Figures 5 and 11), despite the rapid drop-off in the use of the promoted hashtag.
CONCLUSIONS \& FUTURE DIRECTIONS
In this paper, we present a measurement-driven study
of the effects of promoted tweets and trends on Twitter
on the engagement level of users, using a number of ML
and NLP techniques in order to detect relevant tweets
and their sentiments. Our results indicate that the use of
accurate methods for sentiment analysis, and robust filtering for topical content, is crucial. Given this, we then
see that promoted tweets and trends differ considerably
in the form of engagement they produce and the overall
sentiment associated with them. We found that promoted trends lead to higher engagement volumes than
promoted tweets. However, although promoted tweets
obtain less engagement than promoted trends, their engagement forms are often more brand-inclusive.

•features of the event as the location of the event’s peak, its growth and relaxation signatures, contain valuable information and must be used in the analytics. The predictive power of Twitter events in this study was evaluated using two scenarios: in the first scenario, we performed the analysis for the aggregated
Twitter signal considering only tweets’ sentiment; in the second scenario, we cluster Twitter sentiment events based on their shape and then calculate the statistics of successful predictions separately for every event type. The results
of our analysis can be summarized as follows:
- Aggregated Twitter events can be used to predict sales spikes.
- Spikes can be separated into categories based on their shapes (position of
the peak, growth, and relaxation signatures).
- Different spike shapes are differently associated with sales.
- Some spike types have a higher predictive power than the aggregated Twitter signal.

•This shows that Twitter users do not tweet as positively about a promoted trend as they would about a promoted tweet. Instead, a large proportion of tweets relating to a promoted trend contained no emotional words, or if they did, the positive and negative sentiments balanced each other out. They generally contained just the promoted hashtag or generally had an objective, matter-of-fact tone (e.g., - “Get 3G where I live... \#O2WhatWouldYouDo”).Figure 11:\# related engagements

•Social media sites like Facebook and Twitter have emerged
as popular destinations for users to get real-time news about
the world around them. On these sites, users are increasingly
relying on crowdsourced recommendations called Trending
Topics to find important events and breaking news stories.
Typically topics, including key phrases and keywords (e.g., hashtags), are recommended as trending when they exhibit a
sharp spike in their popularity, i.e., their usage by the crowds
suddenly jumps at a particular time (Twitter 2010). Once a
topic is selected as trending, it gets prominently displayed
on social media homepages, thus reaching a large user
population. Additionally, traditional news organizations often pick trending topics and cover them in their news
stories (a practice termed Hashtag journalism), further amplifying their reach. Recognizing their importance,

•Benevenuto, F., Magno, G., Rodrigues, T., and Almeida, V. 2010. “Detecting spammers on Twitter,” in Collaboration, electronic messaging, anti-abuse and spam conference, pp. 12-21. 
Bermingham, A., and Smeaton, A. F. 2011. “On using Twitter to monitor political sentiment and predict election results,” in Proceedings of the Sentiment Analysis where AI meets Psychology Workshop.
Bollen, J., Mao, H., and Zeng, X. 2011. “Twitter mood predicts the stock market,” Journal of Computational Science (2:1), pp. 1-8.
Cha, M., Haddadi, H., Benevenuto, F., and Gummadi, P. K. 2010. “Measuring User Influence in Twitter: The Million Follower Fallacy,” in Proceedings of the Fourth International AAAI Conference on Weblogs and Social Media, pp. 10-17.
Culotta, A. 2013. “Lightweight methods to estimate influenza rates and alcohol sales volume from Twitter messages,” Language resources and evaluation (47:1), pp. 217-238.
Dann, S. 2010. “Twitter content classification,” First Monday (15:12).}

 & {\setlength{\baselineskip}{0.8\baselineskip} Consider using transformer-based models like BERT or GPT for more insightful trend analysis in social media. These models excel in understanding context and semantics, which can enhance sentiment analysis and topic detection. Additionally, exploring techniques like clustering for sentiment events can provide deeper insights into the dynamics of trending topics. \newline \newline \textbf{REFERENCES} \newline \newline •Poddar, D. (2020, April 23). War of the hashtags: Trending new hashtags to override critical topics in social media. \newline •Dacres, S., Haddadi, H., and Purver, M. (2013, December 23). Topic and sentiment analysis on OSNs: A case study of advertising strategies on Twitter. \newline •Kolchyna, O., Souza, T. T. P., Aste, T., and Treleaven, P. C. (2015, August 17). In quest of significance: Identifying types of Twitter sentiment events that predict spikes in sales. \newline •Chakraborty, A., Messias, J., Benevenuto, F., Ghosh, S., Ganguly, N., and Gummadi, K. P. (2017, April 1). Who makes trends? Understanding demographic biases in crowdsourced recommendations. \newline •Dijkman, R., Ipeirotis, P., Aertsen, F., and van Helden, R. (2015, March 16). Using Twitter to predict sales: A case study.} & 0.0059 & 0.25 & 0.8943 \\
\hline
\end{longtable}
\normalsize % Reset to normal size after the table
\end{center}

\newpage

\section*{Appendix E. Sample Output For the Most Enhanced Model For Question 40}
\label{AppendixE}

\begin{center}
\renewcommand{\arraystretch}{1} % Adjust row height
\tiny
\begin{longtable}{|p{1cm}|p{8cm}|p{2.5cm}|p{1.3cm}|p{1.35cm}|p{1.3cm}|}
\hline
%\centering \textbf{Question} & \centering \textbf{Contexts} & \centering \textbf{Answer} & \centering \textbf{Context Relevance}  & \centering \textbf{Faithfulness} & \centering \textbf{Answer Relevance} \\\hline
\textbf{Question}&\textbf{Contexts}&\textbf{Answer}&\textbf{Context Relevance}&\textbf{Faithfulness}&\textbf{Answer Relevance}\\\hline
{\setlength{\baselineskip}{0.8\baselineskip}As a data scientist in sports analytics, I'm clustering player performance metrics. Beyond Gaussian Mixture Models, which advanced clustering method should I consider for improved insights?} & 
{\setlength{\baselineskip}{0.8\baselineskip}Global lifestyle and investment patterns are significantly influenced by socioeconomic changes. The widespread availability of high-speed internet and online social media has enabled people to express their views freely on various scales. Nowadays, a majority of users utilize platforms such as WhatsApp, Twitter, Instagram, etc., to communicate and share comments through tweets, status updates, stories, reels, shorts, etc. The digital revolution has also transformed the financial market, making stocks, mutual funds, and precious metals easily accessible in electronic form, allowing people to make online purchases conveniently from the comfort of their homes. Throughout the last century, economics has achieved unprecedented heights. In the multicurrency system, nearly all governments worldwide strive to manage their country's economic status by boosting GDP, reducing inflation, and controlling the exchange rate between their domestic currency and foreign currency to enhance the lifestyle and income levels of their citizens. Microeconomics focuses on the financial position of entities such as farms and enterprises, guiding decision-making and choice among various options. Stocks, mutual funds, oil, gas, precious metals, goods, services, and products witness fluctuations in their market prices over time. These variations are influenced by factors such as news, natural calamities, political stability, climate, and other dynamic elements. The nature of news has changed considerably, with users now primarily accessing information online through various social media platforms rather than relying solely on traditional printed newspapers. Consequently, social media posts play a crucial role in shaping the financial market, especially when authored by influential leaders, significantly amplifying their impact. In this study, we explore the impact of social media posts, such as tweets, on the financial market using artificial intelligence. We analyze the relationship between tweets and stock prices through the application of Natural Language Processing. This motivation propels us to develop a sophisticated model that can effectively analyze the influence of social media posts, particularly tweets, on financial markets. 1. Beyond Traditional Models and Temporal Sentiment Dynamics: In contrast to existing financial market prediction models employed by investment banks, which predominantly rely on historical price analysis, our model extends its reach by considering the influential voices of market leaders on social media platforms. By incorporating Twitter sentiments, we introduce a state-of-the-art Memory-based Sentiment Analysis, which goes beyond immediate effects by gauging the temporal impact of tweets on financial market prices. Twitter Sentiment: Our model adopts a multistep learning process, beginning with learning to analyze Twitter sentiment. This phase involves the generation of sentiment scores and categorization into positive, negative, or neutral sentiment statuses based on leaders' tweets. This state-of-the-art approach enables the system to discern the emotional tone of leaders' voices in real-time. Additionally, historical sentiment patterns from previous $n$ days are used to capture more enduring effects on market prices. This dual approach enhances our model's predictive accuracy by considering both the immediate and longer-term impacts of social media sentiments on financial markets. Building on sentiment analysis, our model progresses to learn to map tweet sentiments with different financial components. Here, the model establishes a connection between sentiment scores and statuses with dated market prices, effectively synchronizing the ebb and flow of sentiments with market trends. This mapping provides a comprehensive understanding of how social media sentiments correlate with financial components. Mapping: The final phase, learning to predict the price using tweets stock mapping, equips the model with the ability to forecast financial market fluctuations based on the established mapping between social media sentiments and historical prices. This integration allows our model to offer timely and state-of-the-art predictions that consider both traditional market factors and social media dynamics. To enhance interpretability, this essentially teaches a backtraversal mechanism. By analyzing the effects of specific words or sets of words on Twitter that influence market fluctuations, the system gains the ability to identify and highlight the linguistic triggers influencing financial markets. This granular analysis enhances the model's transparency and provides state-of-the-art insights into the relationship between social media sentiments and market movements.}
 & {\setlength{\baselineskip}{0.8\baselineskip} Consider using Memory-based Sentiment Analysis, which captures both immediate and longer-term impacts of tweets on financial markets. This technique allows for the generation of sentiment scores and the mapping of these sentiments to market prices, enhancing predictive accuracy. Additionally, it provides insights into the linguistic triggers that influence market fluctuations, improving model interpretability. \newline \newline \textbf{REFERENCES} \newline \newline •Das, A., Nandi, T., Saha, P., Das, S., Mukherjee, S., Naskar, S. K., and Saha, D. (2024). Effect of Leaders' Voice on Financial Market: An Empirical Deep Learning Expedition on NASDAQ, NSE, and Beyond.} & 1.0 & 0.75 & 0.8618 \\
\hline
\end{longtable}
%\normalsize % Reset to normal size after the table
\end{center}

\newpage
\section*{Declaration of generative AI and AI-assisted technologies in the writing process}
During the preparation of this work the author(s) used ChatGPT in order to improve the readability and language of the manuscript. After using this tool/service, the author(s) reviewed and edited the content as needed and take(s) full responsibility for the content of the published article.

\newpage
%\clearpage

\end{document}